\documentclass[10pt]{article}

\usepackage{array} 
\usepackage{epsfig}
\usepackage{amssymb}
\usepackage{amsmath}                          
\usepackage{graphics,graphpap}
\usepackage{graphicx}
\usepackage{epstopdf}

\renewcommand{\thefootnote}{\fnsymbol{footnote}}

\def\simge{\mathrel{%
   \rlap{\raise 0.511ex \hbox{$>$}}{\lower 0.511ex \hbox{$\sim$}}}}

\def\simle{\mathrel{
   \rlap{\raise 0.511ex \hbox{$<$}}{\lower 0.511ex \hbox{$\sim$}}}}
   
\def\s#1{\setbox0=\hbox{$#1$}%
\rlap{\ifdim\wd0>.7em\kern.22\wd0\else\kern.1\wd0\fi /}#1}

\newcommand{\matel}[3]{\langle #1|#2|#3\rangle}

\newcommand{\dU}{{d_{\cal U}}}
\newcommand{\cU}{{\cal U}}

\begin{document}

\begin{center}
   {\Large \bf \boldmath Unparticles and CP-violation\footnote{Talk contributed to the EPS meeting Manchester July 2007}
}
     \end{center}
     \begin{flushright}\begin{tabular}{l}
IPPP/07/79\\
DCPT/07/158
\end{tabular}
\end{flushright}
 
     {\sc
{\bf Roman Zwicky} \\[0.2cm] 
 {\em IPPP, Department of Physics, Durham University, Durham DH1 3LE, UK}} \\[0.1cm]
e-mail: Roman.Zwicky@durham.ac.uk

\vskip0.4cm

{{\bf Abstract:}
We give a brief summary of the unparticle scenario proposed by Georgi.  The CP-even phase of the propagator is exploited to 
study the CP-asymmetry in $B^+ \to \tau^+ \nu$, which is neither experimentally searched for nor predicted by any other model. Furthermore we show that the novel CP-violation
is consistent with the CPT theorem by identifying the CP-compensating mode in the unparticle sector.}

\setcounter{footnote}{0}
\renewcommand{\thefootnote}{\arabic{footnote}}

\vspace{-0.1cm}
\section*{\normalsize 1.Introduction}
\vspace{-0.1cm}

The possibility of a strongly coupled scale invariant sector, 
weakly coupled to the  Standard Model (SM), 
was advocated by Georgi in \cite{Georgi1,Georgi2}. 
The operators of the scale invariant theory do not describe
single particle excitations but entail a continuous spectrum,
hence the name ``unparticle''. An interesting deconstruction of this spectrum in terms of a particle tower was given in Ref.~\cite{Stephanov}.
According to \cite{Georgi1} at a very high energy scale $M_\cU \gg 1\,{\rm TeV}$  the particle world could be described by the SM and 
a strongly self-coupled  ultraviolet (UV) sector, interacting with each 
other via a 
heavy particle of mass $M_\cU$ and is 
described by the effective non-renormalizable Lagrangian
\begin{equation}
\label{eq:leff}
{\cal L}^{\rm eff} \sim 
\frac{1}{M_\cU^{d_{UV}+(d_{\rm SM}-4) }} O_{\rm SM} O_{UV} 
\stackrel{\rm lower\,\, Energy}{\to} 
\frac{\lambda}{\Lambda_\cU^{d_\cU+(d_{\rm SM}-4)}} O_{\rm SM} O_\cU \,,
\end{equation}
and at some energy $\Lambda_\cU$  
the UV sector flows into a  \emph{strongly} coupled infrared (IR) fixed point where the UV operator undergoes dimensional transmutation
$O_{UV} \to (\Lambda_\cU)^{d_{UV}-\dU} O_\cU$ and 
the coupling indicated above is
$\lambda = c_\cU (\Lambda_\cU / M_\cU)^{d_{UV}+(d_{\rm SM}-4)}$, with $c_\cU$ being a matching coefficient expected to be of
order one. 

From a Lagrangian of the type \eqref{eq:leff}  either 
real \cite{Georgi1} or virtual effects \cite{Georgi2} can be investigated from symmetry
properties and the scaling dimension $\dU$ alone!
The meaning of the real emission of an unparticle  is at present 
unclear or at least model dependent.
Virtual effects are described in a transparent way
within the formalism of perturbative field theory by the propagator, 
which can be constructed from the dispersion relation
\begin{eqnarray}
\label{eq:dispersive}
\Delta_\cU (P^2) \equiv  i \! \! \int_0^\infty d^4 x e^{i p\cdot x}
 \matel{0}{T O_\cU(x) O^\dagger_\cU(0)}{0} 
 = \int_0^\infty \frac{ds}{2\pi} \frac{2{\rm Im}[\Delta_\cU(s)]}{s-P^2-i0} + 
 {\rm s.t.} \,
\end{eqnarray}
It is assumed that  $P^2 \geq 0$ and $P_0 >0$ and s.t.
stands for possible subtraction terms due to non-convergence in 
the UV.  The imaginary part is related to the local matrix element
by the optical theorem
\begin{equation}
\label{eq:im}
2 {\rm Im}[\Delta_\cU(P^2)] = 
|\matel{0}{ O_\cU(0)}{P}|^2 P^{-2} = A_\dU (P^2)^{\dU-2}  \,,
\end{equation}
whose form is dictated by the scaling dimension of $O_\cU$.
The dispersion integral is then elementary \cite{Georgi2,e+e-}
\begin{equation}
\label{eq:prop}
\Delta_\cU(P^2) = \frac{A_\dU}{2 \sin(\dU \pi)}\frac{1}{(-P^2-i0)^{2-\dU}} 
\stackrel{P^2>0}{\to} 
\frac{A_\dU}{2 \sin(\dU \pi)} \frac{e^{-i \dU \pi}}{(P^2)^{2-\dU}} \quad ,
\end{equation}
for appropriate $\dU$ to be discussed below.  
The normalization factor $A_\dU$, which is arbitrary up
to the requirement $\Delta_\cU(P^2) \stackrel{\dU \to 1}{\to} 1/P^2$,
has been chosen to be
$A_\dU = 16 \pi^{5/2} \Gamma(\dU+1/2)
(\Gamma(\dU-1)\Gamma(2 \dU))^{-1}$ \cite{Georgi1}. It is the analytic
continuation of the phase space volume of $\dU$ massless particles based on the observation that the matrix element $\matel{0}{ O_\cU(0)}{P}$ behaves as such.
This led to the statement that an \emph{unparticle looks like a 
non-integral number $\dU$ of massless particles} \cite{Georgi1}.
The propagator \eqref{eq:prop} exposes power like scaling, unlike the logarithmic scaling of the trivial UV fixed point of QCD, and a CP-even phase factor $e^{-i \dU \pi}$, whose consequences have been investigated in many papers and constitutes  the central ingredient to the analysis presented here. The identification 
\begin{equation}
\dU = 1 + \gamma
\end{equation}
of the scaling dimension and the anomalous dimension follows from the
limit to the free propagator.
The lower bound of values for the scaling dimension is  
$\dU \geq 1 + j_L + j_R$, where $j_{L(R)}$ is the Lorentz spin, 
for which the four dimensional conformal group admits unitary 
representations \cite{Mack}. This bound assures the IR convergence
of the dispersion integral \eqref{eq:dispersive}. 
The integral diverges in the UV for $\dU \geq 2$,
but on the other hand the theory is described in the UV by
the non-scale invariant theory of operators $O_{UV}$,
which alters the dispersion integral in the UV.
In principle there is no upper boundary but nevertheless in the
literature most often the values $ 1 < \dU < 2$ are assumed 
without much loss  for the phenomenological analyses.

Scale invariance is expected to be broken at lower energies,
first by the emergence of the weak scale,  
by coupling the unparticle to the Higgs VEV for instance \cite{FRS},
and second in concrete realizations discrete parameters,
such as the number of colours,  might only allow for a near critical behaviour only. The breaking of scale invariance in the IR 
will change the nature of the unparticle as a final state in case
it does not decay beforehand.

The discussion up to now has been mostly formal based on symmetries.
This raises the question of whether there are indeed such theories
in four dimension that flow into a non-trivial IR fixed point.
In Ref.~\cite{Georgi1} the (perturbative) Banks-Zaks \cite{BZ} fixed-point was given as an illustrative example. Walking technicolour
constitute another example, c.f. \cite{physrep} and references therein, where a scale invariant window is needed
in order to suppress flavour changing neutral currents and contributions to the S-parameter. 
Very recently it was shown that half of the supersymmetric gauge theories and around a quarter of the
the non-supersymmetric gauge theories do indeed flow into 
a scale invariant phase \cite{francesco}. 
Furthermore, it was pointed out that in an appropriate limit the so-called higher dimensional (HEIDI) models, c.f. \cite{vBij} and references therein,  assume the
unparticle spectral relation \eqref{eq:im} and therefore reproduce
the unparticle behaviour.
The role of the non-trivial anomalous dimension is mimicked through, possibly fractional, flat extra dimensions accessible to SM singlet fields. It is worth pointing out that these models are renormalizable
for appropriate ranges of the anomalous dimension.

Unparticle like behaviour as in  the propagator \eqref{eq:prop} 
can be observed in well-known theories as well.
For instance the resummation of logarithms due to the emission and absorption of the massless photon in QED leads to an electron propagator  
$S(p) \sim (\s{p}+m)/(p^2-m^2+i0)^{1-\gamma}$ alike
\eqref{eq:prop}  \cite{BS_book} ;
the analogous case of jets in QCD was considered in Ref.~\cite{Neubert}.
Another example is the scale invariant and solvable 
two dimensional Thirring model \cite{Unparty}, 
where the exact propagator $S(x) \sim \s{x}/(-x^2+i0)^{1+ \gamma}$,
which is the two dimensional coordinate space version of the 
unparticle propagator \eqref{eq:prop}. 

\vspace{-0.1cm}
\section*{\normalsize 2. CP-violation in $B^+ \to \tau^+ \nu$}
\vspace{-0.1cm}

\begin{figure}[h]
 \centerline{\includegraphics[width=1.8in]{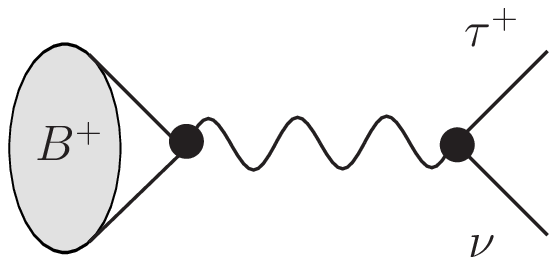}
 \hspace{2cm}
 \includegraphics[width=1.8in]{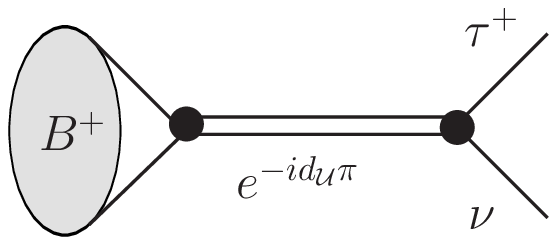}
 }
 \caption{\small (left) SM diagram for $B \to \tau \nu$ (right) unparticle
 diagram with CP odd phase $e^{i\dU\pi}$. 
 }
 \label{fig:btau_feyn}
 \end{figure}

It is  well-known that direct CP-asymmetry occurs if there are
two amplitudes with different weak (CP-odd) and strong (CP-even)
phases, e.g. \cite{PDG}.
In the SM the decay  $B^+ \to \tau^+ \nu$ is mediated 
by the diagram Fig.~\ref{fig:btau_feyn}(left) and receives 
negligable radiative corrections and has therefore no sizable CP asymmetry. Moreover the CP-asymmetry is not even searched 
for in experiment \cite{PDG}!

Coupling a scalar unparticle, in an ad-hoc fashion, to the flavour sector similar to the charged Higgs sector
\begin{eqnarray}
\label{eq:leff1}
{\cal L}^{\rm eff} &=& \frac{\lambda_{S(P)}^{  UD}}{\Lambda_{\cal U}^{d_{\cal U}-1} } ( \bar U (\gamma_5)  D) \, O_{\cal U} \, +\,
 \frac{\lambda_{S(P)}^{\nu_l  l'}}{\Lambda_{\cal U}^{d_{\cal U}-1} } (\bar \nu_l (\gamma_5) l') \, O_{\cal U} \,  +  \, {\rm h.c.} \,,
 \end{eqnarray}
gives rise to the tree-level contribution shown in Fig.~\ref{fig:btau_feyn}(right). The CP-even phase of the propagator and a weak phase  $\lambda^{\rm ub (\tau \nu)}$  
different from either the CKM or the PMNS phase then opens the door to novel
CP-violation,
\begin{eqnarray}
\label{eq:btau_analyze}
 {\cal A}_{\rm CP}(\tau \nu) \equiv
\frac{\Gamma(B^- \to \tau^- \bar \nu) -\Gamma(B^+ \to \tau^+ \nu)}
{\Gamma(B^- \to \tau^- \bar \nu) +\Gamma(B^+ \to \tau^+ \nu)}=  \frac{2 \Delta_{\tau\nu}  \sin(\phi)  \sin(\dU \pi) }{1+  
 2  \Delta_{\tau\nu} \cos(\phi ) \cos(\dU \pi) + 
  \Delta_{\tau\nu}^2}  \,,
  \end{eqnarray}
where $\phi\! =\! \delta \phi_{\rm ub}\!-\! \delta \phi _{\nu l}$ is the non-CKM(PMNS)  weak phase and the ratio of unparticle to SM amplitude is
\begin{eqnarray}
\label{eq:deltatau}
  \Delta_{\tau \nu}  =    
\,
 \rho_{\tau \nu}  \frac{A_\dU}{2 \sin(\dU \pi)} \frac{m_B^2}{m_b m_\tau}    \,
\Big( \frac{m_B^2}{\Lambda_\cU^2}  \Big)^{\dU-1} \,\frac{ (G_F/\sqrt{2})^{-1} }{m_B^2} , \qquad \rho_{\tau \nu} = 
\frac{|\lambda^{\rm ub}\lambda^{\tau \nu }|}{|V_{\rm ub}U_{\tau \nu}|} \,.
\end{eqnarray}
The first factor is the ratio of flavour couplings, the second is 
a normalization factor of the order of one, the third is kinematical,
the fourth measures the relevance of the operator and the fifth 
factor is an enhancement factor $\sqrt{2}(G_F m_B^2)^{-1} \sim 5\cdot 10^3$  which is peculiar to the tree-level
weak unparticle sector \cite{Unparty} and allows for large
effects unlike in other sectors \cite{FRS}.
Two simplifying assumptions were made, first 
$\lambda \equiv \lambda_S = \lambda_P$ and second it was
assumed that the ratio of $\lambda^{l \nu_{l'}}/U_{l \nu_{l'}}$ to be independent of $l'$ which otherwise leads to more complicated
formulae \cite{Unparty} because the neutrino flavour is not observed 
in experiment. The main results, for $\Lambda_\cU = 1\,{\rm TeV}$,  are \cite{Unparty}
\begin{enumerate}
\item The measured branching ratio ${\cal B}(B \to \tau \nu)$
\cite{PDG} allows for maximal CP-violation (suitable  parameters)
\item Up to $\rho_{\tau\nu} \sim 10^{-3}$ parameters   
$(\dU,\phi)$ exist for which ${\cal A}_{CP} \sim 80\%$, due
to the enhancement factor \eqref{eq:deltatau}
\end{enumerate}
CP-violation with unparticles was also studied  in reference 
\cite{unCP}.

\vspace{-0.1cm}
\section*{\normalsize 3. Consistent with CPT?}
\vspace{-0.1cm}
 It is well known that the CPT symmetry enforces the equality of
 the total rate of particle and antiparticle. In fact, the equality already holds for
 the sum of partial rates of particles rescattering into each other, 
 e.g. \cite{wolf}.
\begin{eqnarray}
\label{eq:genCPT}
\sum_{i \in I} \Delta \Gamma(B \to f_i) = 0 \,, \qquad 
\matel{f_i}{S^\dagger}{f_j} \neq 0 \quad i,j \in I \,,
\end{eqnarray}
where $\Delta \Gamma(B \to f)  \equiv \Gamma(B \to f)-
\Gamma(\bar B \to \bar f)$ is the width difference. 
It is not clear which decay channel compensates for
the novel CP asymmetry ${\cal A}_{\rm CP}(\tau \nu) \sim \Delta \Gamma( B^+ \to \tau^+ \nu)$ in the unparticle world.

\begin{figure}[h]
 \centerline{ \includegraphics[width=1.2in]{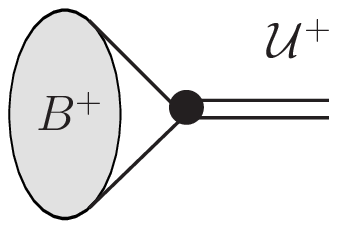}
 \hspace{2cm}
 \includegraphics[width=2in]{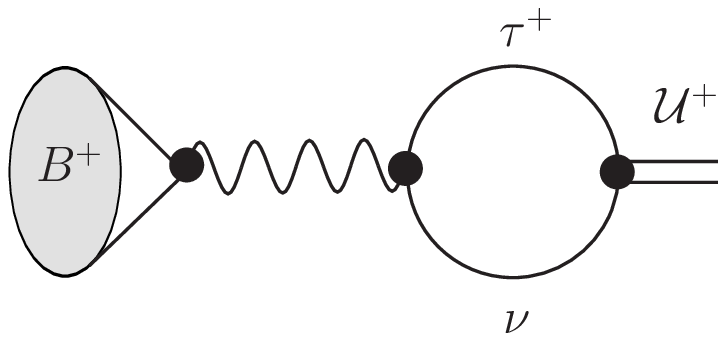}
 }

 \caption{\small $B^+ \to \cU^+$, the double lines denote 
 an unparticle (left)  leading order (right)
 with virtual $\tau\nu$-loop correction.}
 \label{fig:unparticle}
 \end{figure}

In the SM there is no appropriate final state  since
$\tau^+\nu$ is essentially a class on its own. We are led to look in the unparticle sector for a suitable candidate. 
A hint can be gained from counting the weak coupling constants.
The processes $B^+ \to \cal U^+$  with an interference of 
the two amplitudes depicted
in  Fig.~\ref{fig:unparticle} has the same counting in the coupling constants. One amplitude corresponds to a tree decay and the 
other  incorporates a virtual correction due to a fermion loop
of the $\tau$ and the $\nu$.
The process $B^+ \to {\cal U}^+$ is kinematically
allowed since the unparticle has a continuous mass spectrum.
It does not proceed at resonance, but rather behaves like a
multiparticle final state and is a realization of Georgi's observation that
the unparticle field in a final state 
behaves like a non-integral number $\dU$ of
massless particles. 
We refer to Ref.~\cite{Unparty} for further details where
the exact verification of  $\Delta \Gamma(B^+ \to \tau^+ \nu) + 
\Delta \Gamma(B^+ \to {\cal U^+})_{\rm \tau \nu-loop} = 0$ is 
demonstrated explicitly.

 {\bf Conclusions} The unparticle scenario gives rise to spectacular phenomena. For example it permits CP-violation in leptonic decays
 unprecedented so far in other models. We have shown that the CP-violation is consistent with the CPT theorem. Yet a concrete realization 
of the unparticle scenario remains to be worked out, where questions
such as the nature of the real unparticle, the breaking of scale invariance
and the fate of unparticles at low energies can be studied in a
concrete and quantitative way.
 
 \vspace{0.2cm}

{\bf Acknowledgements:} The author is grateful to the organizers of 
the EPS meeting for their dedication and to many colleagues for discussions. Apologies, for all the omitted references.

\vspace{-0.5cm}

\bibliography{referencess}

\end{document}